\documentclass[conference,11pt]{IEEEtran}
\IEEEoverridecommandlockouts
\usepackage[cmex10]{amsmath}
\usepackage{amssymb}
\usepackage{amsthm}
\usepackage[british,english]{babel}
\usepackage{url}
\usepackage{enumerate}
\usepackage{xspace}
\usepackage{balance}

\usepackage{xcolor}
\usepackage{soul}
\definecolor{lightblue}{rgb}{.60,.60,1}

\newcommand{\G}{\textsf{G}\xspace}
\renewcommand{\P}{\textsf{P}\xspace}
\newcommand{\C}{\textsf{C}\xspace}
\newcommand{\V}{\textsf{V}\xspace}

\theoremstyle{theorem}
\newtheorem{thm}{Theorem}
\newtheorem{prop}[thm]{Proposition}

\theoremstyle{definition}
\newtheorem{exmp}[thm]{Example}
\newtheorem{prot}[thm]{Protocol}
\newtheorem{defn}[thm]{Definition}
\newtheorem{assumption}[thm]{Assumption}

\theoremstyle{remark}
\newtheorem{rem}[thm]{Remark}

\DeclareMathOperator{\poly}{poly}

\begin{document}

\title{How to verify computation with a rational network}

\author{
\IEEEauthorblockN{
Sanjay Jain\IEEEauthorrefmark{1},
Prateek Saxena\IEEEauthorrefmark{2},
Frank Stephan\IEEEauthorrefmark{3}, and
Jason Teutsch\IEEEauthorrefmark{4}
}
\IEEEauthorblockA{
School of Computing,
National University of Singapore\thanks{Author order is alphabetical.  P.\ Saxena and J.\ Teutsch research is supported by Singapore Ministry of Education Grant No.\ R-252-000-560-112. S.\ Jain and F.\ Stephan are supported in part by NUS grant No. R252-000-534-112 and R146-000-181-112.  S.\ Jain is also supported by NUS grant No.\ C252-000-087-001.}\\
\IEEEauthorrefmark{1}\url{sanjay@comp.nus.edu.sg},
\IEEEauthorrefmark{2}\url{prateeks@comp.nus.edu.sg}\\
\IEEEauthorrefmark{3}\url{fstephan@comp.nus.edu.sg},
\IEEEauthorrefmark{4}\url{teutsch@comp.nus.edu.sg}
}}

\maketitle

\begin{abstract}
The present paper introduces a practical protocol for provably secure, outsourced computation.  Our protocol minimizes overhead for verification by requiring solutions to withstand an interactive game between a prover and challenger.
For optimization problems, the best or nearly best of all submitted
solutions is expected to be accepted by this approach.
Financial incentives and deposits are used
in order to overcome the problem of fake participants.
\end{abstract}

\section{Introduction}

In the classical Byzantine General's Problem \cite{Lam82}, three parties try to agree on whether to ``attack'' or ``retreat'' by passing messages back and forth.  Here neither ``attack'' nor ``retreat'' represents a semantic truth.  The generals simply want to agree on one plan or the other, and neither plan is \emph{de jure} more correct than the other.  In other consensus situations, such as those involving mathematics, a true answer may in fact exist.  A simple way a group may achieve agreement about a known computational problem is for each party to compute the solution locally.  With local computation, the communication cost between parties is zero and therefore beats any Byzantine consensus protocol in terms of the number of messages sent.
In practice, however, this local computational approach may not be feasible due to its heavy computational burden.

A machine with limited resources may wish to outsource a computational problem to an external system.  Today cloud services exist which provide outsourcing service to businesses and individuals.  These services generally require the individual to trust the cloud which may or may not behave correctly for any reason, including hardware failures or server cheating in order to save computational resources.

One promising approach to ensuring a correct answer from an outsourced computation involves forcing the cloud to provide a short proof which witnesses the correctness of his computation.  Researchers have achieved much progress on this method in recent years \cite{WB15}, however the cryptographic setup costs and computational overhead for state-of-the-art systems make these methods impractical for many applications.  We take a different approach.  In Section~\ref{sec:pcp}, we contrast our approach with this probabilistically checkable proof method.

Let us assume that the individual who wants to outsource his computation offers to pay for a correct answer.  Suppose that a peer-to-peer distributed system of \emph{rational} servers exist who want to solve computational problems in exchange for monetary rewards.  In this paper, we show how to harness the power of economics and consensus in order to incentivize correct computation from such a network with minimal overhead costs.

\subsection{The Verifier's Dilemma}

The Bitcoin network provides evidence that systems can economically incentivize correct computation.  Bitcoin miners correctly solve cryptographic hashing problems in exchange for monetary rewards in the form of electronic currency.  The scope of problems correctly computed on Bitcoin does not, however, extend beyond ``inverting'' SHA2 hashes.  The new cryptocurrency Ethereum \cite{Eth14}, which runs essentially the same consensus protocol as Bitcoin, permits one to achieve an outsourced computation protocol through a series of transactions \cite{LTKS15}.  Anyone can publicly offer a reward for a puzzle, and the first person to announce a correct solution gets the reward.

Under the assumption that Ethereum miners are rational, the Ethereum network may not correctly distribute rewards for solutions whose verification requires more than minimal computational work \cite{LTKS15}.  As in the protocol for Bitcoin, the Ethereum protocol dictates that every miner should check every transaction that appears on the cryptocurrency's \emph{blockchain}, or public ledger.  While Ethereum miners receive rewards for being the first to solve mining puzzles, they do not receive rewards for checking transactions.  Thus miners may skip \emph{heavy} transactions which take a long time to verify in order to get ahead in the mining race.  But whether a rational miner, who is trying to maximize his mining rewards, benefits from verification depends on whether other miners are honestly verifying, building on top of the current blockchain without verifying, or backtracking along the blockchain and extending with a block that doesn't include the heavy transaction.  When a heavy transaction appears on the blockchain, a rational miner doesn't know whether to verify it or not.  This conundrum is known as the \emph{Verifier's Dilemma} \cite{LTKS15}, and Ethereum's \texttt{gas limit} system does not entirely mitigate this problem.  

As a result of the Verifier's Dilemma, the Ethereum blockchain may contain unverified transactions.  An example of a heavy transaction would be a solution to an outsourced computational puzzle which take a long time to verify.  Thus, even though the Ethereum protocol achieves consensus on the blockchain, it may not achieve consensus on the \emph{correct} blockchain.  In particular, assuming rational miners, Ethereum's \emph{consensus computer}, or consensus-based outsourced computation system, yields incorrect results.

\subsection{Contribution of this paper}

We describe a modification to the Ethereum consensus protocol which incentivizes against the Verifier's Dilemma described above.  Under the assumption that computational entities on the network are both:
\begin{itemize}
\item \emph{non-lazy} in the sense that one can convince them to perform computations for the right price and
\item \emph{rational} in the sense that they will try to maximize their individual payoffs from participation,
\end{itemize}
then our modified Ethereum system yields correct results for any feasible function.

In the following discussion, a prover who provides a solution to a puzzle must withstand challenges from other parties on the network.  These challenges result in an authenticated, public transcript of interaction between the prover and challenger which may be communicated on the Ethereum blockchain or through some other means.  We assume throughout this paper that the Ethereum smart contract mechanism can effectively enforce the intended penalty and reward consequences of such transcripts.  The network identifies each party in the protocol by its Ethereum wallet address.

We remark that the Verifier's Dilemma may extend to other cryptocurrencies besides Ethereum, such as Bitcoin, when the number of transactions included becomes large enough that verifying all of them requires significant computational effort.  One can also use the verification game presented in this paper to counteract the Verifier's Dilemma in Bitcoin by appropriate modification to the Bitcoin protocol.  Assuming a modification to the Bitcoin in which miners can challenge blocks with wrong transactions, either by posting ``challenge'' transactions or through some other means and with appropriate rewards for finding errors and penalties for false alarms, one can achieve not only consensus on the Bitcoin blockchain but, in fact, \emph{correct} consensus.

\subsection{Incentives from cryptocurrencies}

Luu, Teutsch, Kulkarni and Saxena \cite{LTKS15} proposed to use
cryptocurrencies, in particular Ethereum, for outsourcing computations.
The network verifies outsourced computations, either deterministically or in a statistical way. The current work 
improves on this by using a stricter verification protocol
by using a verification
game between a prover and a challenger in order to check a proposed solution
to a computation task. The role of the network then becomes to verify that
the players (prover and challenger) stick to the rules of the verification
game.
Provers and challengers have to pay deposits in order to participate.
These deposits are refunded if the players are honest; however, the deposits
of dishonest provers and verifiers are distributed in order to compensate
the honest participants of a verification game for their effort and the
task giver for the additional waiting time.

Under the assumption that at least one participant of the pool of
provers is honest and takes the effort to push his solution
through, the proposed framework leads to the acceptance of a correct
solution at the end. In the case that no one solves the problem, but
at least one challenger is honest and invests the time to challenge
the wrong solutions, no solution
is accepted by the framework and the task giver receives back the
offered prize money. This model resembles the
situation of grand problems in mathematics such as the seven millennium problems
\cite{Ja06}, where one million dollar prize money has been offered for
seven important mathematical problems to be solved and the main
effort of checking the solutions was shifted to the mathematical community.

Consider the case that only a few places in the world
own supercomputers --- such as Quantum computers --- and someone
without direct access to these computers wants to solve a difficult problem,
say factorizing a large integer. He could then outsource the computation
and the verification to such a network. This example also shows why
one should permit challengers who need not contribute a solution: some owners
of traditional supercomputers might not be able to factorize, but are still
able to run the algorithm of Agrawal, Kayal and Saxena \cite{AKS04}
to check whether all factors in a proposed solutions are prime.

\subsection{Systems that outsource computation}

Humans today process more information than they ever have in the past. One typically associates scientific data processing with  ``supercomputers,'' but even your average consumer invokes computationally-intensive processes every time they do a search query on Google, receive news feeds on Facebook, or ask Mathematica to compute Gr\"{o}bner bases.  The utility of mobile devices derives not only from their connectivity to the Internet but also from their connection to more powerful machines.

Cloud computing offers a simple way to outsource computation.  A user who wants to compute something submits a query to the cloud in the form of a program and then waits for the cloud's answer.  In this model, the user must trust that the machine hardware, software, and administrator all function the way she expects them to.  In some situations, one or more of these may not be reasonable assumptions.  Hardware failures occur, software bugs abound, and what's to stop a dishonest cloud that wants to save CPU cycles from spewing a random answer when he knows that the user can't tell the difference from a real one?

Due to the possibility of such errors, one might ask the cloud, in addition to providing the desired computation to also provide a proof of its correctness.  Several recent papers, including \cite{WSHRBW14}, follow this approach.  Since reading a complete proof of correctness might take as much time as simply doing the calculation oneself, the authors employ the machinery of interactive proofs \cite{LFKN92} so that users need not read the entire proof but only sample a few bits to confirm its correctness.  Although constructions have drastically improved efficiency in recent years \cite{WB15}, general-purpose methods at the time of this writing still carry impractically high computational overheads for the cloud server.

\subsection{Our approach}

The approach to outsourced computation described herein achieves provably correct computation through financial incentives rather than proof systems.  We describe a distributed system in which anyone connected to the network can offer a reward for a puzzle, and anyone connected to the network may obtain that reward in exchange for a correct solution to the puzzle.  In order to maximize speed and minimize redundant work, we consider two types of protocols.  In computations where it is easy to randomly guess and check a solution, for example finding a satisfying assignment to a SAT instance, provers can immediately submit their solutions to a puzzle.  This approach gives good speed, and if all the provers guess solutions randomly, the amount of duplicated work is minimal.  In computations which require more verification work, for example multiplying extremely large numbers, the puzzle giver may decide to hold a preliminary bidding round to select a single prover.  This bidding reduces the price the puzzle pay for a correct solution because each prover expects to get paid a fixed amount per unit work, and the bidding process reduces the total work done among all provers.

Our present work introduces a practical, new consensus protocol especially suited for outsourced computation.  The consensus protocol offers a mechanism by which the network can check the correctness of a prover's solution with minimal overhead.  If the given solution has an error, anyone may challenge it (for a reward), then the prover may respond to the challenge, then the challenger may challenge back again, etc.\ in such a way that verifying that the prover and challenger follow the predefined rules of their interaction game requires only a small amount of computational work.  In case the finite interaction game reaches its last possible round, and up to that point no rules have been broken, then the trivial final check is passed to an authority, for example the set of Ethereum miners \cite{Eth14}.  At this point, the authority either accepts the last challenge, in which case the original solution was wrong, or rejected, in which case it was right.

Our consensus protocol compromises between minimal communication complexity and minimal redundant work.  On one extreme, you have the Bitcoin network in which a single announcement of a new block results in immediate consensus among all parties as to which new transactions are acceptable \cite{TS15}. Consensus in Bitcoin, called \emph{Nakamoto consensus} \cite{Nak09b}, achieves low communication complexity in exchange for an extremely high amount of local work from miners.  In the other extreme, Byzantine consensus \cite{Fis83} avoids local work, but
requires a considerable amount of message passing among parties.  Our verification game below requires nontrivial interaction and some local work, but not nearly as much of either as Byzantine or Nakamoto consensus respectively.  Perhaps consequently, it permits more robust verification than the purely Nakamoto-based outsourced computation system proposed in \cite{LTKS15}.

\section{The Verification Game}

The main idea of this paper --- compared to prior work on consensus computing ---
is that there is an explicit game for verifying the solution between
a prover and a challenger. The basic idea is similar to interactive
proof systems, except that both prover and challenger
have limited resources. 

Nodes on the network have four types of computational Tasks:
\begin{enumerate}
\item solving puzzles,

\item challenging puzzle solutions,

\item checking that provers and challengers play the verification game according to the established rules, and

\item doing proof-of-work as required for participation in the blockchain consensus.
\end{enumerate}
The presence of Task~4) implicitly assumes that we embed our verification game system inside of a cryptocurrency such as Ethereum.  One need not have a cryptocurrency system in order to make use of the verification game protocol below, however employing an existing cryptocurrency construction simplifies the implementation.  Task~4) encompasses Task~3) in the following sense.  Nakamoto consensus dictates that miners always mine on the longest blockchain in which all transactions are correct.  Thus miners can vote for or against the validity of a set of transactions by choosing which fork of the blockchain to extend.

Without loss of generality, we may assume that network nodes engage either in some combination of Task~1) and Task~2) or in Task~4), but not a combination of these two sets.  Moreover, we identify Task~3) and Task~4) into a single task.  This disjointness of computational resources may materialize in practice since miners use specialized hardware to achieve Task~3) whereas Task~1) and Task~2) require general-purpose computers.  Therefore one cannot use machines for one of the sets of Tasks on the other.  We summarize our assumed network properties as follows.
\begin{assumption} \label{ass:lolass}
In order to guarantee correctness of the protocol, we assume the following:
\begin{enumerate}[(i)]
\item for each submitted puzzle solution, either:
\begin{enumerate}[a.]
\item at least one node on the network is able and sufficiently non-lazy to both properly check the full solution and raise alerts for mistakes as needed, or
\item a fixed fraction of the nodes of the network are able and sufficiently non-lazy to check a share of the solution and raise alerts for mistakes as needed, and
\end{enumerate}
\item more than half of the nodes who verify the rules of the verification game are \emph{$\epsilon$-rational}~\cite{LTKS15} in the sense that they will do the verification correctly without receiving a reward so long as their computational burden in doing so does not exceed some small threshold $\epsilon$.
\end{enumerate}
\end{assumption}
Whether we choose Assumption~(i)a or Assumption~(i)b above depends on the verification game contracts.  We discuss two possible protocols in the next section, called consensus-competition or consensus-contract

We shall also assume that, in any verification game,
the computations of both the prover and
the challenger run in time polynomial in the size $n$ of the input.
The puzzle giver himself, who wants
the problem to be solved, is sufficiently wealthy
to pay for the computation, but not in possession of the 
necessary computational resources. 
The network (miners) consists of many participants who verify
that both sides in the verification game stick to the rules and that the
winner is determined correctly; however, each single step which they verify
should be of small computational complexity.
\begin{defn}
 In general, a
\emph{$(f,g,h)$-verification game} consists of
\begin{enumerate}[(i)]
\item $f(n)$ rounds on the block chain where
\item  the players can write up to $O[g(n)]$ bits of information in every
round, and
\item a miner in the network spends up to
$\poly[h(n)]$ time for verifying each step. 
\end{enumerate}
Here the time
is measured for random access computation, and thus a miner need not
read the full information provided in a round but
only the relevant parts needed for verification.
\end{defn}
Normally, one would expect $f(n)$ to be small, say constant or
$\log n$. The function 
$h(n)$ is expected to be around $\log n$ or $\poly\log n$,
as this is the time that the miners use for verification.
The function 
$g(n)$ should ideally be around $n$ (say $n \log n$); however, in some
cases, such as in Proposition~\ref{prop:polytimespace},
$g(n)$ may be a polynomial in $n$.
Note that the space $g(n)$ used need not always occupy a blockchain.
One may put these $g(n)$ bits on a publicly readable board where they cannot be altered.

As the above verification game has costs, every participant
in the consensus computation has to place a deposit. This deposit 
of a participant is split
among the hurt parties in the case that the verification game
shows that the participant has been cheating, for example, by submitting
a false solution or challenging a correct solution.

\begin{exmp} \label{exmp:mm}
Matrix multiplication of $n \times n$ matrices over a fixed finite field
$\mathbb F$ has a $(2,n,\log n)$-verification game.

The prover has
solution $C$ for the product of $A$ and $B$.

In the first round, the challenger outputs coordinates $i,j$ such that
\[
c_{i,j} \neq \sum_k a_{i,k} \cdot b_{k,j}
\]
with corresponding evidence consisting of a sequence $d_0,d_1,\ldots,d_n$ of elements in $\mathbb F$
with
\[
d_m = \sum_{k =1}^m a_{i,k} \cdot b_{k,j}.
\] The
network verifies that $i,j$ are coordinates and that
$d_n \neq c_{i,j}$ and that $d_0 = 0$. The challenger loses if this
verification does not go through.

In the second round, the prover tries to defend himself by providing
a $k$ such that $d_k \neq d_{k-1} + a_{i,k} \cdot b_{k,j}$
in $\mathbb F$. If the network can verify this claim then the prover wins, 
otherwise the challenger wins.
\end{exmp}

\begin{rem}
One can combine this idea together with the idea to verify remainders in
order to make a $(O(1),n \log n,\log n)$-verification game
which permits to check whether the product of two $n \times n$ matrices
of $n$-digit numbers gives a claimed $n \times n$ matrix of $2n$-digit
numbers.
\end{rem}

In the example above, as in the subsequent examples below, we do not specify the means by which the challenger discovers the witness for his challenge.  The challenger might be one of the other provers who has lower priority for receiving the reward.  In this case, he has already computed the product himself and can therefore check the correctness of the entries in~$C$ quickly.  The puzzle giver might also
also try to check the prover's answer himself by computing a random handful of entries from~$C$.  If the puzzle giver finds an error, he can proceed with a challenge.  Alternatively, one could modify the verification game for matrix multiplication so as to have the challenges proceed according to Freivald's probabilistic $O(n^2)$-time algorithm for matrix multiplication verification \cite{Fre79,Tha13}.

\begin{exmp}
Consider the problem of finding intersection of two sets
$A$ and $B$ which are a subset of some universal set $U$. 
This problem has a $(2, O(\log n), O(\log n))$-verification game.

Assume that sets $A$, $B$, and the solution $C$
are represented in array form. Furthermore, it is assumed that
equality of two elements from $U$ can be checked fast enough
by the network. The proof can be generalized to various other representations.  Suppose $A$ and $B$ are the input sets.
The prover gives a solution $C$.  Let $n_a, n_b, n_c, n_u$ be the size of the arrays for $A$, $B$, $C$, and $U$ respectively.
\begin{enumerate}[\it Round 1.]
\item
The challenger challenging the solution provides a
counterexample 
by giving either
\begin{enumerate}[(i)]
\item positive indices $i_a \leq n_a$ and $i_b \leq n_b$
(denoting that $A(i_a)=B(i_b)$ but $A(i_a)$ not in $C$),
or
\item a positive index $j_c \leq n_c$ (denoting that $C(j_c) \not
\in A \cap B$).
\end{enumerate}
The miners verify that indeed $i_a \leq n_a, i_b \leq n_b$ and
$A(i_a)=B(i_b)$ in case~(i) and $j_c \leq n_c$ in case~(ii). 
If the above test fails, then challenger loses the game.

\item
To defend himself, the prover gives in case~(i) above
a positive index $i_c \leq n_c$ intended to show that $A(i_a)=C(i_c)$ and in case~(ii)
two positive indices $j_a \leq n_a$ and $j_b \leq n_b$ intended to witness that
$C(j_c)=A(j_a)=B(j_b)$.

The miners verify in case~(i) that
\begin{itemize}
\item $i_c \leq n_c$ and 
\item $A(i_a)=C(i_c)$
\end{itemize}
or in case~(ii) that
\begin{itemize}
\item $j_a \leq n_a$,
\item $j_b \leq n_b$ and
\item $C(j_c)=A(j_a)=B(j_b)$.
\end{itemize}
If the above test fails then the challenger wins the game
else the prover wins the game.
\end{enumerate}

Note the assumption that the comparison of the set elements 
(such as $A(i_a)=B(i_b)$ etc.\ in the protocol above) can be done in 
``small time'' by the network.
In case the representation of the elements is large (say taking $r$ bits),
then we can introduce another round of challenge/verification by prover
or challenger to reduce the time needed by the miners.

In case the prover wants to object that $A(i_a)=B(i_b)$ as provided
by challenger in Round 1, 
he can provide a position where $A(i_a)$ and $B(i_b)$ differ. 
The miners then just verify whether the bits of $A(i_a)$ and $B(i_b)$
at the corresponding positions are equal or not.
Similar method can be used in case challenger wants to object that 
$A(j_a)=C(i_c)$ or $B(j_b)=C(i_c)$.
Note that this would take additional $O(\log r)$ bits of space.
Similar comparison method can also be used for checking the following inequalities:
\begin{align*}
i_a \leq n_a, && j_a \leq n_a,\\
i_b\leq n_b, && j_b \leq n_b,\\
i_c \leq n_c, && j_c \leq n_c.
\end{align*}
\end{exmp}

\begin{exmp}
Consider the problem of sorting $n$ numbers
$A(1), A(2), \ldots, A(n)$ each of at most $r$ bits in non-decreasing 
order.
This has a $(2,\log(n+r),\log n)$-verification game.

The prover provides a solution $f(1), \dotsc, f(n)$, each being
a number of $\log(n)$ bits where this sequence is supposed
to be a sorted permutation of the given items $1, \dotsc, n$.

\begin{enumerate}[\it Round 1.]
\item The challenger provides one of the following challenges:
\begin{enumerate}[(i)]
\item a positive number $j \leq n$ such that
\begin{itemize}
\item $f(j)>n$ or
\item $f(j) <1$;
\end{itemize}
\item two postive numbers $i,j \leq n$ such that
\begin{itemize}
\item  $i \neq j$ and
\item  $f(i)=f(j)$;
\end{itemize}
\item positive numbers $j \leq n$ and $b \leq r$ where
the intention is that $$A[f(j)] > A[f(j+1)]$$ and the least bit where they
differ is $b$-th bit.
\end{enumerate}
In case~(i) or case~(ii), the miners just check whether the claim holds.
If so, then challenger wins the game.
In case~(iii), the miners check whether the $b$-th bit in $A[f(j)]$ is $1$
and $b$-th bit in $A[f(j+1)]$ is $0$. If not, then prover wins the game.
Otherwise, the game proceeds to Round 2.

\item The prover provides a bit $b'$, with $1 \leq b' < b$. The intention
is that $b'<b$ is the bit-position where $A[f(j)]$ and $A[f(j+1)]$ differ.
\end{enumerate}
The miners then just check whether the above claims hold, namely that
the $b'$-th bits of $A[f(j)]$ and $A[f(j+1)]$ are different.
If so, then prover wins the game. Otherwise, challenger wins the game.
\end{exmp}

\begin{exmp}
Computing the greatest common divisor of two $n$-digit numbers
has an $(3,n \log n,\log n)$-verification game.

For the solution, the prover provides the greatest
common divisor $c$ of input numbers $a,b$ plus signed $n$-digit
integers $d,d',e,e'$ such that the following equations hold:
\begin{align*}
c &= d \cdot a + e \cdot b,\\
a &= d' \cdot c,\\
b &=  e' \cdot c.
\end{align*}
Alternatively
one could view this also as a $(4,n \log n,\log n)$-verification
game, where the prover provides $d,d',e,e'$ in the initial round
of verification.

Let $a \% p$ denote the remainder of $a$ by $p$ which is a number
in $\{0,1,\ldots,p-1\}$, even for negative numbers.
\begin{enumerate}[\it Round 1.]
\item The challenger claims that the solution is wrong modulo
a $2 \log n$ digit number $p$; if the solution is indeed false,
such a number $p$ exists by the Chinese Remainder Theorem
and the fact that almost all $n$-digit numbers are smaller than the
product of primes smaller than $n^2$ (which can be
represented by $2 \log n$ digits). The challenger provides
$p$, $a \% p$, $b \% p$, $c \% p$, $d \% p$, $d' \% p$, $e \% p$, $e' \% p$,
and the network checks that
\begin{itemize}
\item these numbers all have at most
$2 \log n$ digits (plus a suitable constant to cover the finitely
many exceptions for small values of $n$), and

\item the equations above hold for these values modulo~$p$.
\end{itemize}

\item The prover, to defend the solution, will select
a $k \in \{a,b,c,d,e,d',e'\}$ and claim that $k \% p$ does not have the
value claimed by the challenger. To substantiate
the claim, the prover provides two lists of numbers:
\begin{itemize}
\item an array of $n$ bits $k_0, \dotsc, k_n$ which supposedly is the binary representation of $k$,
\item and array of numbers $\tilde k_0, \dotsc, \tilde k_n$, each at most $2 \log n$ bits long, which allegedly gives the binary representation of $k_0 \dotsc k_m \mod p$.
\end{itemize}
The network will verify that $\tilde k_n$ differs from what
the challenger had claimed to be $k \% p$, otherwise
the prover loses the game.

\item The challenger selects a number $m$ such that
either
\begin{enumerate}[(i)]
\item $m=0$ and $\tilde k_0 \neq k_0$, or
\item$m>0$ and $\tilde k_m \neq (2 \tilde k_{m-1} + k_m)\% p$.
\end{enumerate}
This can be verified by each miner involved in time $\poly \log n$.
If the complaint is justified, then the challenger wins the game
else the prover wins the game.
\end{enumerate}
\end{exmp}

\begin{rem}
One could replace the checking of the remainder computation --
which involves the usage of $O(n \log n)$ space of the block chain --
by an iterated game between the two players which compare their
computations with some type of interval search for a mistake in the
computation of the remainder. This has $2\log n+O(1)$ rounds and
would result in a $(2\log n+O(1),\log n,\log n)$-verification
game for remainder. In this scenario, it is necessary that the
solution $c$ is accompanied by the numbers $d,d',e,e'$.
\end{rem}

\begin{prop} \label{prop:polytimespace}
Let $\varepsilon$ be any positive rational number, and let $k$ and $h$ be positive constants.
If a computational puzzle can be solved in time $n^k$ and space
$n^h$, then it has a $(O(1),n^{h+\varepsilon},\log n)$-verification
game.
\end{prop}

\IEEEproof
Without loss of generality, the model of computation is a Turing machine.
The basic idea is to check the Turing machine configurations
(instantaneous descriptions). Either the prover or challenger's
computation sequence must be wrong.
Therefore, for finding the cheater, it is enough to 
find a $t$ such that the configuration at time $t$ is
same for both but the configurations are different at time $t+1$, and
then check which transition is valid based on the Turing machine.

When challenging the result of the prover, the challenger wants to
convince the network that the Turing machine computation is different
than the result. So he publishes $i$ (with $i = n^{\varepsilon}$)
configurations spaced out equally over the computation time $n^k$ and named
$y_{0,1}, \dotsc ,y_{0,i}$. The prover can now challenge either that
\begin{enumerate}[(i)]
\item  the input is wrongly encoded in $y_{0,1}$ (with a pointer to the wrongly
coded symbol), or
\item the output encoded
in $y_{0,i}$ is the same as the prover provided, or
\item there is
a $j_0$ such that $y_{0,j_0+1}$ is not obtained from $y_{0,j_0}$
by running the machine $n^{h-\varepsilon}$ steps.
\end{enumerate}

For the first complaint, the network can
compare the two symbols at the two positions and decide whether the
complaint on $y_0$ is justified. For a complaint of the third type,
the challenger can give a pointer to the position of difference
and the network can decide who is right. For a complaint of
$y_{j_0}$ not leading to $y_{j_0+1}$, the process can now be iterated
with the challenger providing a new set
$y_{1,1}, \dotsc ,y_{1,i}$ of configurations spaced out
by $n^{k-2\varepsilon}$ computation steps and again the
prover can either challenge the copying of $y_{0,j_0}$ into $y_{1,1}$
by providing a pointer to the point of difference or challenge
the copying of $y_{0,j_0+1}$ into $y_{1,i}$ by providing
a pointer or challenging the computation from $y_{1,j_1}$ to $y_{1,j_1+1}$
for some $j_1$. In general, for each $u < k/\varepsilon$, provided
that the copying is correct, the game goes from analysing
the configurations $y_{u,1}, \dotsc ,y_{u,i}$ to
$y_{u+1,1}, \dotsc ,y_{u+1,i}$ where $y_{u+1,1} = y_{u,j_u}$
and $y_{u+1,i} = y_{u,j_u+1}$ for some $j_u$ selected by the prover.

In the last round, the computations are each progressing by at most one 
time step
and the prover has to pinpoint the positions at which
the two subsequent configurations did not conform with the
rules of updating the Turing machine configurations in one step.
The difference between the two configurations only is the binary
coding of the head positions, the state and the content of
the cell at the old
head position; the network can check whether, at the position pointed
by prover, the computation conforms with the update rules of the Turing machine.
If so, the challenger wins the game; if not, the prover wins the
game by successfully refuting the alternative computation of the
challenger. Thus the
overall number of rounds is approximately $k/\varepsilon$ which
is a constant, as $h,k,\varepsilon$ are constants.
\qed


\begin{exmp}
While finding a prime factorization in polynomial time needs a quantum computer
(based on any algorithm known up to now),
prime factorization itself has a $(O(1),n^{4+\varepsilon},\log n)$-verification game which can be played by polynomial time players on
both sides of the game. So let $b_1 \dotsb b_m$
be the proposed prime factorization of a given $n\text{-}$digit number $a$.

Recall that by the algorithm of Agrawal, Kayal and Saxena \cite{AKS04}, a
number $b_k$, having up to $n$ bits, is a prime iff there are no numbers
$r,c,d$ satisfying one of the following conditions:
\begin{enumerate}[(i)]
\item $c^d = b_k$ and $1<d \leq 2n$;
\item $b_k = c \cdot d$ and $1<c < n^4$ and $d \neq 1$;
\item $r$ is the least number with
\begin{itemize}
\item $b_k,b_k^2,\ldots,b_k^{n^2} \not\equiv 1 \mod r$,
\item $c < \sqrt{r} \cdot n, \text{ and}$
\item $(x+c)^{b_k} \not\equiv x^{b_k} + c \mod b_k,x^r-1$,
\end{itemize}
where the second modular equivalence is over a polynomial in the formal variable~$x$.
\end{enumerate}
In the game the challenger would either claim
that $b_1 \dotsb b_m$ is not equal to $a$
or that some $b_k$ is not a prime. There is a bound of $O(n^3)$ on
the size of $r$ \cite{LP15} and thus the space used by the
algorithm is $O(n^4)$ which is the space needed to represent
the polynomials $(x+c)^u$ modulo $b_k,x^r-1$ when computing
the power $(x+c)^{b_k}$ modulo $b_k,x^r-1$ fast. The existence
of the verification game now follows from Proposition~\ref{prop:polytimespace}.
\end{exmp}

\begin{defn}
In an optimization problem where one is trying to find a solution maximized on some measure, the \emph{quality} of a solution refers to the magnitude of that measurement.  Thus a high quality solution is closer to optimal than a low quality one.
\end{defn}

\begin{exmp} \label{mlp}
In machine learning, given a sample $(x_1,c_1),\ldots,(x_n,c_n)$
of data items $x_k$ to be classified as $c_k$ by some classifier $f$
(like a perceptron or decision tree) whose parameters are to be learned
by a learning algorithm, one can phrase it as an optimization task.
Here each solution is given of the form
$(\sigma,k_0, \dotsc ,k_n)$ where $\sigma$ is the set of parameters of $f$
and $k_m$ says how many of the first $m$ solutions are classified
correctly; the value $k_n$ is the quality of the solution.
A solution is valid iff
\begin{enumerate}[(i)]
\item $0=k_0 \leq \dotsb \leq k_n$, and 
\item $k_{m+1} = k_m+1$ in the case that
the classifier given by $\sigma$ satisfies $f(x_m) = c_m$
and $k_{m+1} = k_m$ if $f$ satisfies $f(x_m) \neq c_m$.
\end{enumerate}
The network first verifies that $k_0 = 0$ and the
remaining verification game runs in constantly many rounds. The
first round is a challenge of the form $m \in \{0, \dotsc ,n-1\}$
indicating that $k_{m+1}$ is wrongly updated from $k_m$. 
Subsequent rounds then investigate whether the underlying computation
of $f(x_m)$ gives $c_m$, so that one can check whether the challenge
to the update of $k_m$ is justified.

For example, in the case of separation by a 
hyperplane, $\sigma$ will be the coefficients
of the hyperplane and $f(x_m)$ gives the side ($0$ or $1$) of the hyperplane
in which $x_m$ lies in.
In case of classification by
decision trees, $\sigma$ is a particular decision tree. Then, $f(x_m)$ gives the
value of classification as given by the decision tree when the input is $x_m$.
\end{exmp}

\begin{rem}
Similarly, one can also formulate the factorization game as an optimization
task. In this case, any solution of an $n$-bit number $a$ to be factorized
consists of a list $(m,b_1,\dotsc ,b_k)$ such that
\begin{itemize}
\item $m=k$,
\item $2 \leq b_1 \leq \dotsb \leq b_k$, and
\item $a = b_1\dotsb b_k$.
\end{itemize}
 Any solution
not obeying these rules is invalid and the larger the $m$ the
better the solution. Note that the number $m$ can be written down
using $\log(n)$ bits (as $m \leq n$) and one can verify solutions
in LOGSPACE. Thus the game has, for every $\varepsilon > 0$,
a $(O(1),n^\varepsilon,\log n)$-verification game.
\end{rem}

\section{Frameworks of Consensus-Competition and Consensus-Contract} \label{sec:cccc}

These verification games can be used to put together a 
framework to carry out consensus computations where the bulk of the
work is done by the task giver who wants a task to be solved,
the provers who hand in possible solutions to be checked and the
challengers who are incentivized to check the handed-in solutions
in order to eliminate faulty ones.

We present below two verification game variants, which we call consensus-competition (Protocol~\ref{prot:comp}) and consensus-contract (Protocol~\ref{prot:cont}).  While the consensus-competition protocol has the advantage that one can verify some puzzles more quickly and that the work of the challenger may be reduced, however we will focus primarily on the consensus-contract protocol for its ease of implementation.   In a consensus-competition, challengers race to check a prover's solution using probabilistic or any ad-hoc methods, and the first one to find an error wins a prize.  In a consensus-contract, the challengers collectively decide ahead of time, by bidding or otherwise, who should check the given solution first.  When puzzles do not lend themselves to guess-and-check error finding, consensus-contracts have the advantage that they reduce the amount of redundant work done by the network.

\begin{prot} \label{prot:comp}
The framework of a \emph{consensus-competition} is the following. For covering
the network costs, it is assumed that all participants (task-givers,
provers and challengers) provide some fees to the network for the costs
incurred by the network. These costs are constants and are left out from the 
following description. The deposits mentioned below do
not include these fees.

\begin{enumerate}[(a)]
\item The task giver $T$ specifies a task to be solved, an algorithm
for the verification process and the time permitted for phases (b) and (d)
listed below (where the time of (d) might be a formula depending on the number
$m$ of solutions), and the run time permitted for each step in the
verification game. 
The task giver also specifies the amount of the prize
to be awarded for the solution and also the minimum deposit
to be paid by each prover and challenger.
Additionally, for optimization problems, the task giver may provide
a minimal quality of the solution needed to enter the competition.

\item The provers provide encrypted pairs $(S,q,d)$ where $S$ is the solution
and $d$ the amount of deposit which they offer, and in case of optimization
problems, $q$ is the quality of the solution.
Note that $d$ and $q$ must be above the minimum as specified in (a).
Higher amount of deposit is used to get higher priority as mentioned below.
Both numbers $d,q$ are assumed to have only small number of digits
so that the network working bound of the protocol is met.

\item Each prover decrypts his solution and deposit in a verifiable way
and the network sorts the pairs $(S,q,d)$ -- solution, quality (of
optimization problems) and deposit -- in an order such that higher
quality solutions are before lower quality solutions and, for equal
quality solutions, higher deposits go before lower deposits. The network
furthermore eliminates solutions not meeting the minimum deposit or
quality criterion.
Let $S_1,S_2,\ldots,S_m$ be the solutions in the above order.

\item In the next round, each prover
and each member of the network
can challenge solutions. The priority among the challenges are again given
by encrypted deposits, where one deposit works for all challenges by the
same user. A prover is allowed to top up its deposit on the solution in order
to get a higher priority for challenges. The network orders the challenges
according to the amount of the deposit given by them.

\item The network maintains list of pairs of solutions and challenges in
lexicographic order (first all pairs involving $S_1$, then those involving
$S_2$ and so on). In each round one processes the least pair $(S_i,C_j)$
in the above order where neither $S_i$ nor $C_j$ have been disqualified so far,
until the list is exhausted. For each such pair $(S_i,C_j)$, the
players $S_i$ and $C_j$ play the verification game and the network
decides who wins and who loses the game.
The loser loses his deposit which is split into half / half between the
winner and $T$; furthermore, the loser will now be disqualified
from further games.
This process terminates when for all pairs $(S_i,C_j)$ in the list,
either the game has been decided or one of the parties has been
disqualified.

\item If no solution survives, then $T$ gets his prize refunded.
If one or several solutions survive, then the least $i$ such that $S_i$
survives receives the prize.
Furthermore, the deposits of all surviving provers and challengers
get refunded to the party who paid the deposit.
\end{enumerate}
\end{prot}

\begin{prot} \label{prot:cont}
A \emph{consensus-contract} is a contract between a task-giver $T$
and a client $S$ where $T$ gets the solution of $S$ checked
in a way similar to a consensus-competition. $S$ has to pay a
deposit for his solution to be checked. If within
the agreed time-frame no challengers successfully
challenge the solution, then the deposit is refunded to $S$ together
with the prize for solving the problem; the deposits of the challengers
are split between task-giver and prover in order to compensate them
for the delays and work caused by the checking. If some challenger
successfully challenges the solution $S$, then the deposit of $S$
is split between task-giver and challenger and the prize is recycled
for the next contractor willing to take up the task. Here each challenge
is handled in the same way as before by a verification game and
challengers are put in an order based on the deposits that they
provide.
\end{prot}

It can be easily verified that among the solutions, the prover $S_i$
providing a correct solution wins (with $i$ being minimal, in case of
several correct solutions),
provided that $T$ gives enough time
for provers and challengers to compute the winning strategies for the
corresponding verification games: Each $S_j$ with $j<i$ will be defeated
by a sufficiently well-equipped challenger and $S_i$ will defend its
solutions successfully to all challenges. So the task should be solved
successfully provided that $T$ gives enough time to compute solutions,
make the moves in the verification games and provides enough incentive
for an honest and sufficiently capable player to participate in the
competition for the solution.
This is essentially the content of the following theorem.
A similar result can also be shown for optimization tasks.

\begin{thm}
Assume that the following conditions hold:
\begin{itemize}
\item $A$ is a task which has exactly one output for every input;
\item The task giver provides a prize money which is at least two times
   the cost of computing the solution and the network fees paid by the provers,
   for every sufficiently powerful prover of the network;
\item The deposit required is at least four times
   the maximum of following costs: 
\begin{itemize}
\item the network fees to be downpaid when submitting
   a solution or challenge, and 
\item the expected local computation costs for a
   sufficiently powerful prover or challenger to do the computations
   to carry out all rounds of the verification game;
\end{itemize}
\item The task giver provides enough time for phase (b) so that a member of
   the network with sufficient computation power can solve the problem within
   the specified time;
\item The task giver provides enough time for phase (d) so that a member of
   the network with sufficient computation power can compare each solution
   with its own solution and, in the case that it differs, submit a
   challenge;
\item The task giver provides enough time for each step in the verification
   game so that the provers and challengers with sufficient computation power 
   can run their winning strategy for the case that their solution or
   challenge, respectively, is correct.
\end{itemize}
Then, in the case that one prover with sufficient computation power
participates in the game, this prover is incentivized to submit a correct
solution and to win the prize money and one such prover will eventually
win the prize money. Furthermore, the deposits motivate challengers to
detect false solutions and provers to defend correct solutions and thus
deter provers and challenges from putting up false solutions or challenges.
\end{thm}

\IEEEproof
Note that if the provers and challengers have enough time for the verification
game, then they can challenge any wrong solution. The costs involved
for them is the network cost paid and the cost of computation.
If they win the challenge, then they get twice the above costs as payback
(as the remaining deposit of the loser goes to the task giver).
Thus, there is enough incentive for the challengers to challenge wrong
solutions.

If the prover $S$ has enough time to compute a solution, then its
costs are just the network fees and the cost of computation. If the
prize money is double the above costs, then there is enough incentive
for $S$ to enter the contest. As mentioned above, $S$ also has
enough time/incentive to challenge a higher priority wrong solution.
Thus, either $S$ wins, or only correct solutions with priority higher
than $S$ survive. In either case a correct prover wins the prize money.
\qed

Note that if the provers/challengers solve the problem not just based
on reward/deposit gained but also based on expected chance that they
win the reward in case of presence of large number of provers, then
the above incentive structure need to be appropriately updated.
For example, if the number of expected correct solutions posted 
is $5$, then for a prover expected return is 1/5-th of the reward.
Thus, the reward needs to be at least 5 times the costs. However,
as it is not possible to estimate the number of provers who will attempt
the problem, we have used twice the costs as enough incentive.

\begin{rem}
A consensus-contract has the advantage that contract-taker 
knows that the prize will be awarded
to him, provided that a correct solution is given (and thus
the verification goes through). 
The task-giver has the advantage that the prize money does not need to be
too high: the provers will only try to solve if their expected 
award is more than the work done ---
for example in the consensus competition case,
if the expected number of correct
submitted solutions is three, then each prover knows that he gets the
prize only with probability $1/3$ and therefore he would only work
if either the prize money is at least three times the expected costs
or if he has the means to bypass the other correct solutions through
a higher deposit. 
For tasks where
many users spend a small amount of effort to solve the problem and where
solving the problem is a game of chance -- like guessing next
week's lottery numbers or, in a computational setting, finding the solutions
of some satisfiability problem -- the consensus-competition might be
the more adequate way to go. On the other hand, where the task is 
deterministic and effort needed is well defined, just large, consensus-contract
is the more adequate way to go.
\end{rem}

\begin{rem}
For the choice of the parameters in a consensus-competition, the following
ideas might be helpful as a guideline.
\begin{itemize}
\item The encrypted phase (b) is there to avoid the case that
  some users read the solutions of the other users and
  cooperate with miners in order to get their solution into the blockchain
  with priority (timestamp) ahead of the others. Furthermore, paying a
  deposit is like having some confidence that the solution will survive
  a verification game; hence solutions with higher deposits are
  given priority over solutions with lower deposit.
\item Deposits might not deincentivize malicious players on the network if
  they have enough money. They might, however, compensate the persons
  involved for the work they did (such as provers playing a verification game
  to justify the solution, challengers playing a game to eliminate fraudulent
  provers and the task-giver suffering a delay from ongoing fraudulent
  behavior of provers and challengers).
\item The task giver has to select the amount of the prize and the
  minimum deposits for provers and challengers and the time given for
  the competition carefully. Too low minimum deposits attract cheaters who
  try to get the prize unjustifiedly; too high minimum deposits might
  prevent honest provers or challengers from participating due to the
  lack of funds. Too high prize money or too long solving time might
  attract too many solutions, which then makes the whole process lengthy
  and results in many verification games to be played. Too short solving
  time might prevent any honest prover from being able to submit a solution.
\item Similarly the task giver has to decide about the time-frame in
  which players have to move in the verification game or other parts
  of the process. If players stop to move they must get disqualified; however, 
  in order to avoid honest players getting disqualified due to the
  time-complexity of computing the right move, the time-out must be
  sufficient generous. On the other hand, a too high time-out might slow down
  the whole computation and decision process.
\item Malicious players might, under various names, put up several 
  solutions. A player providing a correct solution is allowed to use
  his single deposit to challenge all other solutions in order to prevent 
  that he goes bankrupt when challenging a multitude of malicious 
  incorrect solutions of higher priority; similarly a challenger who does not
  provide a solution does also pay only one deposit, though he might
  challenge all solutions.
\item In order to avoid that the system is abused by a player making
  multiple false challenges, all challenges and the solution of a player
  are thrown out when he loses a verification game --- honest players
  are supposed to win them under all circumstances. Thus each
  run of the verification game is paid by the deposit of the loser
  (a prover or a challenger).
\item Players might circumvent the rule that false solutions or challenges
  throw them out of the competition by using different identities; however,
  then they have to also pay multiple deposits, and thus this does not pay
  off financially.
\item The verification game will make the best solution win phase (e) of
  the protocol for the consensus computation provided that there is a
  good solution and that its prover challenges all false solutions which
  pretend to be better.
\item Phase (d) allows additional challengers to enter the game.
  For example, a large factorization problem might only be solvable
  using quantum computers, while primality testing algorithms 
  \cite{AKS04} permit sufficiently powerful traditional
  supercomputers to check for the primality of the factors of these 
  large numbers;
  hence they are allowed to enter the game. As indicated, the verification games
  then use only $O(n^{4+\varepsilon})$ space on the block chain entries
  and this makes it possible for $T$ to get the factorization task 
  verified through the network rather than by using its own limited resources.
  Furthermore, phase (d) also allows $T$ to challenge 
  incorrect solutions and avoid 
  the situation where it is forced to award a prize unjustifiedly.
\end{itemize}
\end{rem}

\section{A verification protocol with incentives} \label{sec:eskimo}

We assume that individuals on the network are rational, in the sense of Nash equilibrium, and wish to exchange computational resources for financial rewards.  We want to incentivize correct computations.  At a high level, our protocol is as follows.  
\begin{prot} \label{prot:gpcv}
A solution to a puzzle is presented, followed by challenges.
\begin{enumerate}[\it Step 1.]
\item A puzzle giver \G presents a puzzle.

\item A prover \P proposes a solution $S$ to \G's puzzle.

\item A small set of randomly selected challengers $\C_1, \C_2, \dotsc$ checks whether $S$ is correct.
\begin{enumerate}
\item If there are no objections from the challengers, then
\begin{itemize}
\item the network \emph{accepts}~$S$,
\item \P pays \G for his solution, and
\item each of the challengers receives a nominal reward for participating.
\end{itemize}

\item If some $C_i$ reports a mistake in $S$, then:
\begin{itemize}
\item  $C_i$ checks the solution $S$ by playing the verification game with $\P$, and
\item the loser of this verification game pays a large penalty to the winner.
\end{itemize}
\item Next:
\begin{itemize}
\item In case this game continues until the last possible round, the community \V determines the winner by verifying the correctness of the final response in the game.
\item In case $\V$ determines that $\C_i$ made a false alarm, after $\C_i$ pays a penalty we repeat Step~3.
\item Otherwise $\C_i$ receives her reward from for correctly raising an alarm, and the network \emph{rejects} $S$.
\end{itemize}
\end{enumerate}
\end{enumerate}
\end{prot}

We now give more details.

 \subsection{Step 1: Presenting the puzzle}
  
The puzzle giver posts his puzzle in a public place.  The puzzle may be a known problem, such as checking whether all the transactions in a Bitcoin block are valid, or a puzzle that an individual independently wants to solve.  External to the system we describe, $\G$ must commit both his puzzle and its prize to this system so that in the end the system can either reward or penalize the prover for his solution.  For instance, the problem and its reward could be committed to some public blockchain.  $\G$'s puzzle is converted into a verification game either manually or via Proposition~\ref{prop:polytimespace}.

\subsection{Step 2: Committing a solution}

In this step, the network selects a candidate solution~$S$ for the puzzle.  The prover~\P commits a significant amount of capital to an escrow which, according to some contract, he gets back later iff his solution~$S$ turns out to be correct.  The prover wants to be sure that no one else can steal his answer and claim a reward for his solution, and the community wants to be sure that the prover can't change his solution once it has been committed.

The paper~\cite{LTKS15} describes a commitment scheme in which a prover's solutions might be securely broadcast and selected using the Ethereum blockchain.  First, everyone races to solve the puzzle, and then an Ethereum blockchain lottery elects a prover from among the first solutions to appear.  The designated prover then hashes his solution onto the blockchain, thereby committing his answer, and once the blockchain decides that the prover $\P$ has been elected, he then commits his solution in plaintext to the blockchain along with a smart contract carrying his deposit.

\subsection{Selecting a subcommittee}

Next we select a random subcommittee of challengers to check $\P$'s solution~$S$.  The number of challengers selected should be large enough that at least one among them is honest enough to check the solution~$S$ in exchange for the potentially large reward for finding a mistake, and the subcommittee size is not too large so as to be prohibitively expensive.  The expense here derives from the fact that each member of the subcommittee must be compensated financially for his computational effort in checking, so larger subcommittees are more expensive.  The rewards for challengers should be at least enough to compensate for checking effort in case the solution turns out to be correct, and \emph{significantly} more in case the challenger finds an error.  The ``significantly'' part is needed in order to incentivize the challenger to actually do the checking, particularly in the case where the challenger believes that his chances of finding an error are small.

We select the subcommittee of challengers via a lottery.  When the solution is announced, everyone who is interested in participating in the subcommittee does a proof-of-work.  Ideally the proof-of-work puzzle should be \emph{sequential} rather than \emph{parallelizable}, as defined in \cite{KMS15}, so that each party, regardless of computational power, takes the same amount of time to complete the proof-of-work task.  Then everyone with a minimum amount of computational power has a chance to win the race and therefore gets a lottery ticket.  Sequential proof-of-work differs from the parallelizable work used in Bitcoin and other cryptocurrencies An ideal candidate for sequential proof-of-work has yet to appear \cite{KMS15}, nevertheless, but we need only imagine that everyone who can demonstrate a minimum amount of computational power should have a chance to be selected.  Finally we randomly sample a subset of those who successfully completed the proof-of-work task.  If the proof-of-work task had several solutions, we could, for example, elect those members to subcommittee whose hash of their solution ends in some prescribed sequence of digits.

In addition to submitting a proof-of-work, each candidate for the subcommittee must deposit some money into an escrow account.  This money will be used to impose a penalty in case the candidate is selected for the subcommittee and subsequently raises a false alarm (or an incorrect check).  We omit precise details of the subcommittee construction as we anticipate embedding this game inside of another forthcoming system which can securely appoint the subcommittee \cite{But15,GHK13}.

\subsection{Playing the verification game}

Each challenger privately checks the solution~$S$.  The checking round ends as soon as one of the following things happens:
\begin{itemize}
\item some challenger $\C_i$ announces a mistake in $S$, or


\item the predetermined time limit for challenging has been reached.
\end{itemize}
Due to the non-laziness assumption, at least one challenger will respond within the given time limit.  In the latter case, each $\C_j$ gets paid for verification in proportion to the \emph{gas limit}, or predetermined CPU cycles, for the problem.  If, on the other hand, some challenger $\C_i$ announces a mistake in $\C_i$, then $\C_i$ and $\P$ play a verification game.  In this case, $\C_i$ will either receive a substantial reward for detecting a mistake or face a substantial penalty for raising a false alarm.

The community vote $\V$ only gets invoked if the final round of the verification game occurs.  Hence if the challengers agree that $\P$'s solution is correct or if $\C_i$ and $\P$ can settle their verification game amicably, then the community need not do any work.  In this way we minimize the computational burden on the community which, as in Bitcoin, we expect does a small amount of verification work for free.

\subsection{Correctness and efficiency}

We assume that challengers in the network are rational, and that there are sufficiently many of them that are non-lazy so that at least one non-lazy challenger belongs to each randomly selected subcommittee, thereby satisfying Assumption~\ref{ass:lolass}-(i)a and Assumption~\ref{ass:lolass}-(ii).  In Section~\ref{sec:cccc}, we discussed two flavors of verification games based on either consensus-competition or consensus-contract protocols.  Consensus-competitions cost less because the challengers only sample the output and do not necessarily check the whole thing.  Consensus-contracts, on the other hand, cost more because challengers are expected to replicate the prover's entire work.  For simplicity, we will discuss only consensus-contracts here.   In general, challengers have incentive to be both non-lazy and verify honestly in this protocol because they gain large prizes for discovering errors, and they are discouraged from reporting errors falsely by monetary penalties.  

For example, suppose that the challenger is supposed to check whether a matrix multiplication is correct, and suppose that a consensus-competition compensates challengers for the expected work required to find an error assuming 10\% of the entries are wrong, and that the prize for finding an error during this check is 50 times more than the compensation for checking.  If the challenger believes that the actual chances of finding a mistake in the prover's solution are only 1/5000, then the rational challenger views this 50$\times$ prize as inadequate and therefore lacks incentive to do the check.  This discrepancy between the challenger's belief regarding the probability of finding an error and his actual probability of finding an error exists whether we employ a consensus-contract or a consensus-competition.

The difference between these methods is that in a consensus-competition, the challenger might honestly do his verification, find no mistakes in his samples, and yet not receive a reward because the overall submitted solution is actually incorrect.  This outcome unfairly penalizes the challenger, and so a rational challenger might not accept such a contract in the first place.  Whether or not a challenger will accept such a contract for matrix multiplication depends on what the challenger believes is the true probability of finding an error.  Since the challenger does not \emph{a priori} know how likely an anonymous prover is to make an error, no rational argument can say whether or not the challenger should agree to the consensus-competition.  In contrast, a rational challenger should accept an adequately compensated consensus-contract, and if the reward for finding a bug suffices to convince the challenger to be non-lazy, then the challenger will do the intended work and give the correct answer.  We incentivize all of the challengers on the subcommittee to respond by only paying compensation once sufficiently many challengers have responded.

It is not unreasonable to expect that the community will do a negligible about of work for free.  We see this commonly today in various distributed systems.  The challengers on the network may function also as verifiers.

\section{Practical considerations}

\subsection{Outsourced computation} \label{sec:pcp}

Mobile, scientific, and other various big data applications often require local machines to outsource computations to an external cloud computer.  How can one be sure that the computation done by a cloud is correct given only limited local resources?  Computations can contain errors due to hardware failures, software errors, or even laziness to expend resources on the part of the cloud.  The past five years have witnessed a rising interest in secure, outsourced computation protocols in which the cloud service provides, in addition to the computation itself, some certificate of the computation's correctness which is easy for the client to verify \cite{WB15}.  Ideally, time required to verify the computation should be small compared to the time required to simply execute the computation directly, while the cloud's extra overhead in producing the certificate should be minimal.  The techniques applied thus far involve PCP-type encodings and SNARK-like cryptographic protocols.

One can use the protocol in Section~\ref{sec:eskimo} to outsource computation.  The puzzle giver $\G$ puts forth the problem and then some prover $\P$ puts forth a solution.  The economics of the verification system ensure that $\G$ receives a correct answer.  In Walfish and Blumberg's survey article ``Verifying Computations without Reexecuting Them'' \cite{WB15}, the authors argue against replicating computations as a means to achieving secure computation.  Potential solutions to this problem, they say, suffer from either systemic errors which appear in repetitions across cloud platforms, wrongly assume that auditors can detect rare errors, or rely on assuming that a particular piece of hardware will execute correctly.  Since anyone can participate in the verification network, we make no assumptions about how a particular cloud sets up its hardware, and we employ powerful economic forces to not only detect, but reduce, errors.

Our present system offers some advantages over the PCP/SNARK outsourced verification systems to date.  Firstly, it doesn't use PCPs or SNARKS.  Our protocol is simple in the sense that we do not require any complicated encoding schemes or cryptographic protocols.  Secondly, our protocol can be used to correctly compute any polynomial-time functions, whereas the PCP-style systems proposed thus far generally operate effectively on smaller classes of functions \cite{WB15}.  Finally, as we argue below, our system has low computational overheads to the point of being practical.

Our system has negligible setup cost.  In contrast, many of the proof-based systems to-date, including Zaatar, Pinocchio, Ginger, and TinyRAM, require one to run thousands of instances of a single function before breaking even on the overall verification time on a $128 \times 128$ matrix multiplication puzzle \cite{WB15}.  Some systems, only run on restricted circuits or try to amortize setup costs.  CMT, Allspice, and Thaler effectively slash verification time with negligible overhead costs.  The prover overhead for Thaler in particular, under matrix multiplication, has the same order of magnitude worker overhead as the procotol we have presented here.

In many practical cases, a user might wish to reuse input, rather than reuse a function, in order to execute several analyses on the same data.  In such cases, proof-based systems with overhead generally become impractical on the verification side.  Our incentive-based system, however handles data-analysis cases well as there is no overhead in switching between functions.

In general, the prover side of these proof-based systems is the bottleneck which prevents them from being practical.  With the exception of Thaler, all of the systems benchmarked in \cite{WB15} require a factor of $10^5$ worker overhead for the prover when performing a $128 \times 128$ matrix multiplication.  Thus the cloud processing power being used should be significantly more than $100,000$ times the verifier could do on his own hardware in order for outsourcing to make any sense.  Matrix multiplication, in our protocol, can be done with a 3-round verification game plus the verification time from the community (see Example~\ref{exmp:mm}).  So if we assume that the communication time for posting the problem and electing a prover to give a solution takes no more time than solving the problem itself, we can conclude that the prover overhead for matrix multiplication, in terms of both computational time and effort, is generously at most a factor of 5.  This is much cheaper than the above systems in terms of computation and hence in terms of dollars as well.  In general, an arbitrary polynomial-time function can be done with $\log n$ overhead (Proposition~\ref{prop:polytimespace}).

\subsection{Outsourced verification}

Cryptocurrencies like Bitcoin maintain a public ledger, called a blockchain, which includes all transactions that take place over the network.  Miners on the network are responsible for maintaining the integrity of these transactions through two steps:
\begin{enumerate}
\item they select which transactions to include in the blockchain, and

\item they verify that blockchain transactions are valid.
\end{enumerate}
As the Verifier's Dilemma in \cite{LTKS15} illustrates, miners do not always verify transactions correctly.  We can use our protocol in Section~\ref{sec:eskimo} as a means to allow miners to offload their verification burden, thereby relieving them of the second task.  Here each miner plays the role of the Puzzle Giver in Protocol~\ref{prot:gpcv} who provides a block of transactions to be verified, along with a financial contract which guarantees funding for the network to perform the verification.  The source of these funds may come either from the miners themselves, or in the case of resource-heavy transactions, from the party who posts the transactions.  The system now places a $\surd$ or a $\times$ next to each transaction (or block of transactions) depending on whether it is correct or incorrect.

Since miners can rely on the $\surd$ and $\times$ marks for correctness, they do not need to verify these transactions themselves.  Therefore this verification protocol reduces the amount of redundant work done on the cryptocurrency.  Only a few challengers in Protocol~\ref{prot:gpcv} check each block, hence the total amount of work spent in verification across both the cryptocurrency and verification network is less than what would be spent if each miner on the cryptocurrency verified each transaction for himself.  In this way, a verification network can offer positive ecological impact.


\section{Conclusion}

By restricting Ethereum to transactions which require little verification time, we achieve a consensus computer which achieves correct outsourced computation results for feasible functions when the network's computational entities are rational.  The verification game described herein permits us to verify, through financial incentives, interaction, and Nakamoto consensus, any polynomial-time puzzle with only $\poly \log$ effort.  We show how to embed this game into real systems and argue that the computational overhead of the implementation is less than the overhead for state-of-the-art for probabilistically-checkable-proof systems.
\balance
\bibliographystyle{IEEEtranS}
\bibliography{cryptogame}

\begin{thebibliography}{10}
\providecommand{\url}[1]{#1}
\csname url@samestyle\endcsname
\providecommand{\newblock}{\relax}
\providecommand{\bibinfo}[2]{#2}
\providecommand{\BIBentrySTDinterwordspacing}{\spaceskip=0pt\relax}
\providecommand{\BIBentryALTinterwordstretchfactor}{4}
\providecommand{\BIBentryALTinterwordspacing}{\spaceskip=\fontdimen2\font plus
\BIBentryALTinterwordstretchfactor\fontdimen3\font minus
  \fontdimen4\font\relax}
\providecommand{\BIBforeignlanguage}[2]{{%
\expandafter\ifx\csname l@#1\endcsname\relax
\typeout{** WARNING: IEEEtranS.bst: No hyphenation pattern has been}%
\typeout{** loaded for the language `#1'. Using the pattern for}%
\typeout{** the default language instead.}%
\else
\language=\csname l@#1\endcsname
\fi
#2}}
\providecommand{\BIBdecl}{\relax}
\BIBdecl

\bibitem{AKS04}
\BIBentryALTinterwordspacing
M.~Agrawal, N.~Kayal, and N.~Saxena, ``\BIBforeignlanguage{English}{{PRIMES} is
  in {P}},'' \emph{\BIBforeignlanguage{English}{Annals of Mathematics}}, vol.
  160, no.~2, pp. 781--793, 2004. [Online]. Available:
  \url{http://www.jstor.org/stable/3597229}
\BIBentrySTDinterwordspacing

\bibitem{But15}
V.~Buterin, ``Notes on scalable blockchain protocols,''
  \url{http://github.com/vbuterin/scalability_paper/raw/master/scalability.pdf},
  2015.

\bibitem{Fis83}
\BIBentryALTinterwordspacing
M.~J. Fischer, ``\BIBforeignlanguage{English}{The consensus problem in
  unreliable distributed systems (a brief survey)},'' in
  \emph{\BIBforeignlanguage{English}{Foundations of Computation Theory}}, ser.
  Lecture Notes in Computer Science, M.~Karpinski, Ed.\hskip 1em plus 0.5em
  minus 0.4em\relax Springer Berlin Heidelberg, 1983, vol. 158, pp. 127--140.
  [Online]. Available: \url{http://dx.doi.org/10.1007/3-540-12689-9_99}
\BIBentrySTDinterwordspacing

\bibitem{Eth14}
E.~Foundation, ``{Ethereum's white paper},''
  \url{https://github.com/ethereum/wiki/wiki/White-Paper}, 2014.

\bibitem{Fre79}
\BIBentryALTinterwordspacing
R.~Freivalds, ``Fast probabilistic algorithms,'' in \emph{Mathematical
  Foundations of Computer Science 1979}, ser. Lecture Notes in Computer
  Science.\hskip 1em plus 0.5em minus 0.4em\relax Springer Berlin Heidelberg,
  1979, vol.~74, pp. 57--69. [Online]. Available:
  \url{http://dx.doi.org/10.1007/3-540-09526-8_5}
\BIBentrySTDinterwordspacing

\bibitem{GHK13}
\BIBentryALTinterwordspacing
R.~Guerraoui, F.~Huc, and A.-M. Kermarrec, ``Highly dynamic distributed
  computing with byzantine failures,'' in \emph{Proceedings of the 2013 ACM
  Symposium on Principles of Distributed Computing}, ser. PODC '13.\hskip 1em
  plus 0.5em minus 0.4em\relax New York, NY, USA: ACM, 2013, pp. 176--183.
  [Online]. Available: \url{http://doi.acm.org/10.1145/2484239.2484263}
\BIBentrySTDinterwordspacing

\bibitem{Ja06}
A.~M. Jaffe, ``The millennium grand challenge in mathematics,'' \emph{Notices
  of the AMS}, vol.~53, no.~6, 2006.

\bibitem{KMS15}
J.~Katz, A.~Miller, and E.~Shi, ``Pseudonymous broadmast and secure computation
  from cryptographic puzzles,'' \url{http://eprint.iacr.org/2014/857.pdf}.

\bibitem{Lam82}
\BIBentryALTinterwordspacing
L.~Lamport, R.~Shostak, and M.~Pease, ``The byzantine generals problem,''
  \emph{ACM Transactions on Programming Languages and Systems}, vol.~4, no.~3,
  pp. 382--401, Jul. 1982. [Online]. Available:
  \url{http://doi.acm.org/10.1145/357172.357176}
\BIBentrySTDinterwordspacing

\bibitem{LP15}
J.~Lenstra, H.W. and C.~Pomerance, ``Primality testing with gaussian periods,''
  2015, \url{https://math.dartmouth.edu/~carlp/aks06-2015.pdf}.

\bibitem{LFKN92}
\BIBentryALTinterwordspacing
C.~Lund, L.~Fortnow, H.~Karloff, and N.~Nisan, ``Algebraic methods for
  interactive proof systems,'' \emph{Journal of the ACM}, vol.~39, no.~4, pp.
  859--868, Oct. 1992. [Online]. Available:
  \url{http://doi.acm.org/10.1145/146585.146605}
\BIBentrySTDinterwordspacing

\bibitem{LTKS15}
L.~Luu, J.~Teutsch, R.~Kulkarni, and P.~Saxena, ``Demystifying incentives in
  the consensus computer,'' in \emph{Proceedings of the 22nd ACM SIGSAC
  Conference on Computer and Communications Security (CCS 2015)}.\hskip 1em
  plus 0.5em minus 0.4em\relax New York, NY, USA: ACM, 2015, pp. 706--719.

\bibitem{Nak09b}
S.~Nakamoto, ``Bitcoin: A peer-to-peer electronic cash system,''
  \url{http://bitcoin.org/bitcoin.pdf}.

\bibitem{Tha13}
\BIBentryALTinterwordspacing
J.~Thaler, ``\BIBforeignlanguage{English}{Time-optimal interactive proofs for
  circuit evaluation},'' in \emph{\BIBforeignlanguage{English}{Advances in
  Cryptology -- CRYPTO 2013}}, ser. Lecture Notes in Computer Science,
  R.~Canetti and J.~A. Garay, Eds.\hskip 1em plus 0.5em minus 0.4em\relax
  Springer Berlin Heidelberg, 2013, vol. 8043, pp. 71--89. [Online]. Available:
  \url{http://dx.doi.org/10.1007/978-3-642-40084-1_5}
\BIBentrySTDinterwordspacing

\bibitem{TS15}
F.~Tschorsch and B.~Scheuermann, ``Bitcoin and beyond: A technical survey on
  decentralized digital currencies,''
  \url{http://eprint.iacr.org/2015/464.pdf}.

\bibitem{WSHRBW14}
R.~S. Wahby, S.~Setty, M.~Howald, Z.~Ren, A.~J. Blumberg, and M.~Walfish,
  ``Efficient ram and control flow in verifiable outsourced computation,''
  Cryptology ePrint Archive, Report 2014/674, 2014,
  \url{http://eprint.iacr.org/}.

\bibitem{WB15}
\BIBentryALTinterwordspacing
M.~Walfish and A.~J. Blumberg, ``Verifying computations without reexecuting
  them,'' \emph{Communications of the ACM}, vol.~58, no.~2, pp. 74--84, Jan.
  2015. [Online]. Available: \url{http://doi.acm.org/10.1145/2641562}
\BIBentrySTDinterwordspacing

\end{thebibliography}

\end{document}